\documentclass[12pt,epsf,amssymb]{article}
\usepackage{epsfig}
\usepackage{amsmath}
\usepackage{amssymb}
\usepackage{axodraw}

\makeatletter

\usepackage{verbatim}

\newcommand{\bea}{\begin{eqnarray}}
\newcommand{\eea}{\end{eqnarray}}
\newcommand{\beq}{\begin{equation}}
\newcommand{\beqns}{\begin{eqnarray*}}
\newcommand{\eeqns}{\end{eqnarray*}}
\newcommand{\eeq}{\end{equation}}

\newcommand{\pdir}{p\kern -5.2pt\raise 0.2ex\hbox {/}}
\newcommand{\vdir}{v\kern -5.75pt\raise 0.15ex\hbox {/}}
\newcommand{\kdir}{k\kern -5.75pt\raise 0.15ex\hbox {/}}
\newcommand{\epsdir}{\epsilon\kern -5.0pt\raise 0.15ex\hbox {/}}
\newcommand{\bvdir}{\bar{v}\kern -5.75pt\raise 0.15ex\hbox {/}}
\newcommand{\Ddir}{D\kern -7.75pt\raise 0.20ex\hbox {/}}
\newcommand{\Adir}{A\kern -7.75pt\raise 0.20ex\hbox {/}}
\newcommand{\ldir}{l\kern -5.0pt\raise 0.2ex\hbox{/}}
\newcommand{\varepsdir}{\varepsilon\kern -5.5pt\raise 0.15ex\hbox{/}}

\newcommand{\msb}{\overline{\rm{MS}}}

\newcommand{\Ga}{\not\!\!\!\ \Gamma}

\newcommand{\ra}{\rightarrow}

\newcommand{\gmu}{\gamma_{\mu}}

\newcommand{\m}{\mu}

\newcommand{\bsb}{\overline{B_s}}

\newcommand{\lgl}{\langle}
\newcommand{\rgl}{\rangle}
\newcommand{\aspi}{\frac{\alpha_s}{4\pi}}

\def\elematrice#1#2#3{\lgl#1|#2|#3\rgl}

\def\Journal#1#2#3#4{{#1} {\bf #2}, #3 (#4)}

\def\NPB{{\em Nucl. Phys.} B}
\def\PLB{{\em Phys. Lett.}  B}

\def\PRep{\em Phys. Rep.}
\def\PRD{{\em Phys. Rev.} D}

\makeatother

\begin{document}

\thispagestyle{empty}

\begin{flushright}
\begin{tabular}{l}
{\tt DESY 07-057}\\
{\tt SFB/CPP-07-16}\\
\end{tabular}
\end{flushright}
\vskip 1cm\par
\begin{center}
{\par\centering \textbf{\Large Lattice renormalisation}}\\
\vskip .45cm\par
{\par\centering \textbf{\Large of ${\cal O}(a)$ improved heavy-light operators}}\\
\vskip .45cm\par
\vskip 0.9cm\par
{\par\centering 
  Beno\^\i t~Blossier$^a$}
{\par\centering \vskip 0.5 cm\par}
{\par\centering \textsl{$^a$ 
DESY, Platanenallee 6, D-15738 Zeuthen, Germany.}}
\end{center}

\begin{abstract}
\noindent The analytical expressions and the numerical values of the renormalisation constants
of ${\cal O}(a)$ improved static-light currents are given at one-loop order of perturbation theory 
in the framework of Heavy Quark 
Effective Theory: the static quark is described by the HYP action and the light quark is 
described either with the Clover or the Neuberger action. These factors are relevant
to extract from a lattice computation the decay constants $f_B$, $f_{B_S}$ and the set of 
bag parameters $B_i$ associated with $B-\overline{B}$ mixing phenomenology in the Standard
Model and beyond.
\end{abstract}
\vskip 0.3 cm

\setcounter{page}{1}
\setcounter{equation}{0}

\renewcommand{\thefootnote}{\arabic{footnote}}
\setcounter{footnote}{0}

\section{Introduction}

The extraction of important quantities like $V_{ub}$ or $|V_{ts}/V_{td}|$ needs the non
perturbative calculation of the hadronic form factors that encode the long-distance physics. For example
the $B$ meson decay constant $f_B$ has to be precisely known to determine the exclusive $V_{ub}$
from $B \to \tau \bar{\nu}$ \cite{vubtaunu}. The detection of physics beyond the
Standard Model in the $B_s$, $\bsb$ system is hopeless if the theoretical uncertainty on the bag
parameter $B_{B_s}$ associated with the $B_s-\bsb$ mixing amplitude in the Standard Model is not
reduced \cite{bsbmixing}. The most satisfying approach to compute such form factors is lattice
QCD, as it is only based on first principles of quantum field theory. However, discretisation
effects induce important systematic errors if $am_Q\geq 1$, where $a$ is the lattice spacing and
$m_Q$ is the heavy quark mass. The extrapolation to the continuum limit of physical quantities involving such
heavy quarks is difficult, 
unless the calculation is done on a very fine lattice (e.g. $a\sim 0.02$ fm), 
which is not possible for the moment because of the too high cost in computation time, or employing the
Relativistic Heavy Quark action \cite{RHQ} with properly tuned parameters \cite{Christ} (see \cite{CPPACS-RHQ}
for a recent application of this approach). A way around 
this problem is the use of Heavy Quark Effective Theory (HQET) \cite{HQET} in which all 
degrees of freedom of ${\cal O}(m_Q)$ are integrated in Wilson coefficients, where 
$m_Q \gg \Lambda_{QCD}$. This approach is attractive because the continuum
limit exists and results are independent of regularisation. A strategy to renormalise non perturbatively the
theory has been proposed and tested for a simple case~\cite{alphaREN}. A drawback of the standard Eichten-Hill 
action \cite{eichten} is the rapid growth of the statistical noise on the correlation functions $C(x_0)$ at $x_0 \sim 1$ fm, 
making difficult the extraction of hadronic quantities. A method to reduce UV fluctuations is the use of HYP links 
\cite{hasen} to build
the Wilson line of the static propagator; it has been found that this strategy improves significantly 
the signal/noise ratio \cite{Alpha2005}. In this paper we give the analytical expressions and the numerical results of the 
renormalisation
constants of static-light bilinear and four-fermion operators at one-loop of perturbation theory when the
static quark is described by the HYP action and the light quark is described by the ${\cal O}(a)$ improved Clover
action or the Neuberger action \cite{Neuberger}; in the latter case the extraction of the bag parameters $B_i$ is much 
safer theoretically 
because there is no mixing among dimension 6 four-fermion operators of different chirality. This work is an extension
to smeared static quark actions of similar computations done with the Eichten-Hill action and with the Clover 
\cite{gimenez,Sachrajda} and Neuberger actions \cite{becirevic} respectively. The first of these two new results might
be used by the authors of \cite{fBscalUKQCD} to give the final number of the ${\rm N_f}=2$ P wave static-light 
decay constant computed with the HYP action.
The paper is organised as follows:
in Section~2 we will present results obtained by using the tree-level improved 
static-light operators and in Section~3 we will give renormalisation 
constants of four-fermion operators, leaving the presentation of the numerical result of the bag 
parameter $B_{B_s}$ to a future paper.

\section{Tree level improved static-light current}

\noindent A well known approach to reduce the cut-off dependence of matrix elements computed on the lattice is to improve 
the Wilson light quark action by adding an ${\cal O}(a)$ term which is irrelevant in the continuum limit, for 
example the Sheikholeslami-Wohlert Clover
one \cite{SW}. One needs also to improve the inserted operators: in the literature, authors defined 
rotated 
fields $\psi' \equiv \left(1 - a\frac{r}{2} \not \!\! D \right) \psi$ \cite{rotatepert}.
We will choose $r=1$ for the rest of the paper. In principle one could also rotate the static field but it has been shown that it is not 
necessary in the computation of ${\cal O}(a)$ improved
on shell matrix elements at tree level \cite{pittori}. A tree-level, the improved bilinear static-light operator will then 
read

\beq
O^I_\Gamma\equiv \bar{h} \Gamma \psi' = \bar{h} \Gamma \psi -\frac{a}{2} \bar{h} \Gamma  \not \!\! D \psi\, ,
\eeq
where $\Gamma$ is any Dirac matrix and we choose the symmetric definition of the covariant derivative 
$D_\mu \psi(x) = \frac{U_\mu(x) \psi(x+\hat{\mu} ) - U^\dag(x-\hat\mu ) \psi(x - \hat\mu )}{2a}$. The 
static quark action reads

\beq
S^{\rm HQET} = \sum_n h^\dag(n) \left[ h(n)-V^{\dag, {\rm HYP}}_4(n-\hat{4})h(n-\hat{4})\right] 
+ a\,\delta m\, h^\dag(n) h(n),
\eeq
where $V_4$ is a HYP-smeared link in time direction and $\delta m$ is a counter-term introduced to cancel the 
linear divergent part of the static quark self-energy \cite{eichten}. The light quark action reads

\beq
S^{\rm Clover} = S^W - a^4 c_{SW} \sum_{n,\mu,\nu} \left[ i g \frac{a}{4} \bar{\psi}(n) \sigma_{\mu\nu} P_{\mu\nu}
\psi(n)\right],
\eeq
where $P_{\mu\nu}$ is the discretised strength tensor. The Sheikholeslami-Wohlert coefficient $c_{SW}$ can be fixed at its tree 
level value $c^{\rm tree}_{SW}=1$ to be consistent with a one-loop calculation in perturbation theory. We collect in 
Table \ref{tab1} the Feynman rules
which are used. We follow the notations of \cite{Capitani} - \cite{Lee} in the rest of the paper and we 
summarise them in Appendix~A.

\begin{table}
\flushleft{
\begin{tabular}{|c|c|}
\hline
{\small static quark propagator}& {\small $a(1-e^{-ip_4a}+\epsilon)^{-1}$}\\
{\small vertex $V^a_{\mu,hhg}(p,p')$}&{\small $-ig_0 T^a h_{4\mu} 
e^{-i(p_4+p'_4)\frac{a}{2}}$}\\
{\small vertex $V^{ab}_{\mu\nu,hhgg}(p,p')$}&{\small $-\frac{1}{2}ag^2_0 h_{4\mu}h_{4\nu}
\{T^a,T^b\}e^{-i(p_4+p'_4)\frac{a}{2}}$}\\
{\small light quark propagator}&{\small $a\left(i \gamma\cdot\bar{p} + am + \frac{1}{2}\hat{p}^2\right)^{-1}$}\\
{\small vertex $V^a_{\mu,qqg}(p,p')$}&{\small $-igT^a(\gmu \cos a(p+p')_\mu-i\sin a(p+p')_\mu)$}\\
{\small vertex $V^{ab}_{\mu\nu,qqgg}(p,p')$}&{\small $\frac{iag^2_0\delta_{\mu\nu}}{2}\left\{T^a,T^b\right\}
\left( \gmu \sin a(p+p')_\mu + i \cos a(p+p')_\mu \right)$}\\
{\small improved vertex $V^I_{\mu,qqg}$}&{\small $-g_0 T^a \frac{r}{2}\left[\sum_\nu \sigma_{\mu\nu}(\sin 
a(p-p')_\nu \cos \frac{a}{2} (p-p')_\mu\right]$}\\
{\small static-light bilinear current $O_{\Gamma_1}$}&{\small $\Gamma_1$}\\
{\small improved static-light bilinear current $O^I_{\Gamma_1,qq}$}&{\small $-\frac{i}{2} \Gamma_1 \Ga$}\\
{\small improved static-light bilinear current $O^I_{\Gamma_1,qqg}$}&{\small $-\frac{a\,ig_0\,r}{2}\Gamma_1 
\gamma_\mu \cos a(p+p')_\mu$}\\
{\small gluon propagator in the Feynman gauge}&{\small $a^2\delta_{\mu\nu}\delta^{ab}(2W + a^2\lambda^2)^{-1}$}\\
\hline
\end{tabular}
}
\caption{\label{tab1} Feynman rules.}
\end{table}
\noindent Note that $p'$ and $p$ are the in-going and out-going fermion momenta, respectively.
We also introduce an infrared regulator $\lambda$ for the gluon propagator. We symmetrize the vertex 
$V^{ab}_{\mu\nu,hhgg}$ by introducing the anti-commutator of $SU(3)$ generators, normalized by a 
factor~$\frac 1 2$. 

\noindent At one loop of perturbation theory, a bare matrix element regularised and renormalised in a continuum 
scheme - for example in the Dimension 
Regularisation (DR) and in the $\overline{\mbox{MS}}$ scheme - is written
generically in terms of its tree level part

\beq
\lgl O(p,\mu) \rgl^{\mbox{DR},\overline{\mbox{MS}}} = \left[1 + \frac{\alpha^{\overline{\mbox{MS}}}
_s(\mu)}{4\pi} \left( \gamma \ln\left(\frac{\mu^2}{p^2}\right) + C_{\rm DR}\right)
\right] \lgl O(p) \rgl^{\rm tree}\,,
\eeq
where $\gamma$ is the ${\cal O}(g^2)$ coefficient of the anomalous dimension of the operator. The same bare matrix element 
regularised on the lattice reads 

\beq
\lgl O(p,a) \rgl^{\rm lat} = \left[1 + \frac{\alpha_{s0} (a)}{4\pi} \left(\gamma \ln (a^2p^2) + C_{\rm lat}\right)\right] 
\lgl O(p) \rgl^{\rm tree}\, +{\cal O}(a)\,.
\eeq
At this level of perturbation theory one can identify $\alpha^{\overline{\mbox{MS}}}
_s(\mu)$ with the bare coupling $\alpha_{s0} (a)$. One can then write that 

\bea\nonumber
\lgl O \rgl^{\mbox{DR},\overline{\mbox{MS}}} &=& \left[1 - \frac{\alpha_{s0}(a)}{4\pi} 
\left(\gamma \ln a^2\mu^2 + C_{\rm lat} - C_{\rm DR}\right) \right]
\lgl O \rgl^{\rm lat}\, +{\cal O}(a)\\
\label{matching}
&\equiv&Z(a\mu) \lgl O \rgl^{\rm lat}\,+{\cal O}(a)\,.
\eea
The matching constant between the matrix element renormalised at the scale $\mu=a^{-1}$ in the continuum and the 
bare matrix element 
regularised on the lattice is then given by $C_{\rm lat} - C_{\rm DR}$. In the following we will be
concerned with the static-light currents and discuss $C_{\rm lat}$.

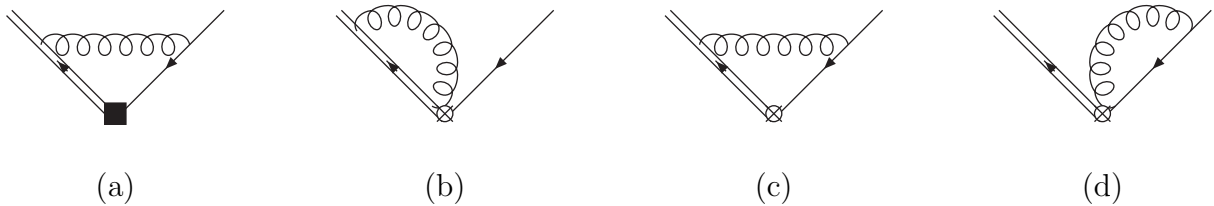
\begin{figure}[t]
\begin{center}
\begin{tabular}{ccccccc}

\begin{picture}(0,30)(0,0)
\Line(1,1)(-39,41)
\ArrowLine(41,41)(0,0)
\ArrowLine(-18,18)(-22,22)
\Gluon(-28,28)(28,28){4}{6}
\Line(-1,-1)(-41,39)
\CBox(-4,-2)(4,6){Black}{Black}
\end{picture}

&\rule[0cm]{3.0cm}{0cm}&
\begin{picture}(0,30)(0,0)
\Line(1,1)(-39,41)
\ArrowLine(41,41)(0,0)
\ArrowLine(-18,18)(-22,22)
\GlueArc(-19,19)(20,-45,135){4}{7}
\Line(-1,-1)(-41,39)
\CCirc(0,2){3}{Black}{White}
\Line(-3,-1)(3,5)
\Line(-3,5)(3,-1)
\end{picture}

&\rule[0cm]{3.0cm}{0cm}&
\begin{picture}(0,30)(0,0)
\Line(1,1)(-39,41)
\ArrowLine(41,41)(0,0)
\ArrowLine(-18,18)(-22,22)
\Line(-1,-1)(-41,39)
\Gluon(-28,28)(28,28){4}{6}
\CCirc(0,2){3}{Black}{White}
\Line(-3,-1)(3,5)
\Line(-3,5)(3,-1)
\end{picture}
&\rule[0cm]{3.0cm}{0cm}&
\begin{picture}(0,30)(0,0)
\Line(1,1)(-39,41)
\ArrowLine(41,41)(0,0)
\ArrowLine(-18,18)(-22,22)
\GlueArc(19,19)(20,45,225){4}{7}
\Line(-1,-1)(-41,39)
\CCirc(0,2){3}{Black}{White}
\Line(-3,-1)(3,5)
\Line(-3,5)(3,-1)
\end{picture}

\\
&&&&&&\\
(a)&&(b)&&(c)&&(d)\\
\end{tabular}
\end{center}
\caption{\label{fig1} Diagrams giving the 1 loop correction to the ${\cal O}(a)$ improved 
static-light current with the 
${\cal O}(a)$ improved light quark action.}
\end{figure}

\noindent Let us consider the bare hadronic matrix element regularised on the lattice 
$\elematrice{H_2}{O^I_\Gamma}{H_1}^{\rm lat}$ where $H_1$ contains the light quark $q$ and $H_2$ contains the 
static quark $h$. It is computed from the ratio
\bea\nonumber
R(t,t_1,t_2)&=&{\cal Z}_1{\cal Z}_2 \frac{C^{(3)}_{J_1,O^I_\Gamma,J_2}(p,p',t,t_1,t_2)}
{C^{(2)}_{J_1}(\vec{p},t_1) C^{(2)}_{J_2}(\vec{p}',t_2-t)}
\eea
where
\bea\nonumber
C^{(2)}_{J_i}(\vec{p},t)&=&\sum_{\vec x} e^{i\vec{p}\cdot \vec{x}} \lgl J_i(t,\vec{x}) J^\dag_i(0) \rgl
\eea
is a 2-point correlation function, $J_i$ is an interpolating field of the hadron state $H_i$ containing either the static quark field 
$h$ or the light quark field $q$,
\bea\nonumber
C^{(3)}_{J_1,O_\Gamma,J_2}(\vec{p},\vec{p}',t,t_1,t_2)&=&\sum_{\vec{x},\vec{y}} e^{i(\vec{p}\cdot\vec{x} 
- \vec{p}'\cdot\vec{y})}\lgl J_2(t_2,\vec{y}) O^I_\Gamma(t) J^\dag_1(t_1,\vec{x})\rgl\,
\eea
is a 3-point correlation function in which the operator $O^I_\Gamma$ is inserted at time $t$.\\ 
Eventually 
${\cal Z}_i = \elematrice{H^{(0)}_i}{J^\dag_i}{0}$, where $H^{(0)}_i$ is the hadron ground state containing either 
the static 
quark $h$ or the light quark $q$. As usual we determine $\elematrice{H^{(0)}_2}{O^I_\Gamma}{H^{(0)}_1}^{\rm lat}$ in the 
interval of $t$ where $R(t,t_1,t_2)$ 
is constant (i.e. ground states are safely isolated). As the spectator quark does not play any role in the renormalisation 
of $O^I_\Gamma$, one may relate $\elematrice{H^{(0)}_2}{O^I_\Gamma}{H^{(0)}_1}^{\rm lat}$ to 
$\elematrice{\bar{h}(p')}{O^I_\Gamma}{q(p)}^{\rm lat}$. That is why it is justified to
compute the matching constants between the currents renormalised in a continuum scheme and the bare currents 
regularised on the lattice by considering 
the matrix elements of quarks\footnote{The renormalisation constants computed in the MOM scheme are actually extracted 
numerically on the lattice by considering such matrix elements \cite{momscheme}.}, which are the only states 
appropriate to do perturbative 
calculations. We stress that the mass counter-term $\delta m$ is cancelled in $R$: thus we will not consider it in our
one loop computations.\\
At this order of perturbation theory, $\elematrice{\bar{h}(p')}{O^I_\Gamma}{q(p)}^{\rm lat}$ is given by

\bea\nonumber
\elematrice{\bar{h}(p')}{O^I_\Gamma}{q(p)}^{\rm lat}&=&\sqrt{Z_{2h}}\sqrt{Z_{2l}} \left\{1+\aspi C_F 
\left[
-\ln (a^2\lambda^2) + d_1 +  n -(l+m)\right.\right.\\
\nonumber
&&\left.\left.\hspace{2cm} +G (d_2 + h - q - 2 d^I) \right]\right\}
\elematrice{\bar{h}(p')}{O_\Gamma}{q(p)}^{\rm tree}\\
&\equiv& Z_{\rm lat}\elematrice{\bar{h}(p')}{O_\Gamma}{q(p)}^{\rm tree}\, ,
\eea
where
\beq\nonumber
\gamma_0 \Gamma \gamma_0 = G \Gamma, \quad \sqrt{Z_{2h}}=1 + \aspi C_F\left(\frac{e}{2} - \ln (a^2\lambda^2)\right),
\quad \sqrt{Z_{2h}}=1+ \aspi C_F\left(\frac{f+f^I+\ln (a^2\lambda^2)}{2}\right)\,;
\eeq
$d_1 + (d_2 - d^I) G$, $hG$, $n  - (q + d^I)G$ and $-(l+m)$ are 
contributions given by the 1PI vertex diagrams shown in Figure \ref{fig1} and $Z_{2h,l}$ come from the quark self
energies. Finally the expression of $C_{\rm lat}$ reads
\beq
C_{\rm lat} = \frac{e+f+f^I}{2} + d_1 +  n -(l+m) +G (d_2 + h - q - 2 d^I).
\eeq
We have collected the numerical values of the various constants in Table \ref{tab2} for the
HYP parameter sets $\alpha_i=0$ (corresponding to standard Eichten-Hill action), $\alpha_1=1.0$, 
$\alpha_2=\alpha_3=0$ (corresponding to APE blocking \cite{APE}), $\alpha_1=0.75, \alpha_2=0.6, 
\alpha_3 = 0.3$ (HYP1) and $\alpha_1=1.0, \alpha_2=1.0, \alpha_3 = 0.5$ (HYP2); their analytical expression is 
written in Appendix~B, 
while we have collected $C_{\rm lat}$ in terms of $\alpha_i$ for axial and scalar static-light currents 
in Table 
\ref{tab3}. For the first set of $\alpha_i$ our results agree with \cite{gimenez,Sachrajda}.
\begin{table}[b]
\begin{center}
\begin{tabular}{ccc}
\begin{tabular}{|c|c|c|c|c|}
\hline
$\alpha_i$&0&APE&HYP1&HYP2\\
$e$&24.48&3.17&2.52&-3.62\\
$d_1$&5.46&4.98&4.99&4.72\\
$d_2$&-7.22&-3.33&-3.70&-1.87\\
$d^I$&-4.14&-2.79&-2.80&-1.99\\
$h$&-9.98&-3.40&-4.43&-1.95\\
$n$&0.73&-2.33&-1.80&-2.88\\
$q$&-2.02&-0.61&-0.78&-0.19\\
\hline
\end{tabular}
&&
\begin{tabular}{|c|c|}
\hline
$f$&13.35\\
$f^I$&-3.63\\
$l$&-3.42\\
$m$&7.35\\
\hline
\end{tabular}\\
\end{tabular}
\end{center}
\caption{\label{tab2} Numerical values of contributions to the correction at one loop of perturbation theory
of the ${\cal O}(a)$ improved static-light current regularised on the lattice to its tree level expression;
$f$, $f^I$, $l$ and $m$ are extracted from \cite{pittori} whereas $e$ was computed in \cite{blossier}.}
\end{table}

\begin{table}[b]
\begin{center}
\begin{tabular}{|c|c|c|c|c|}
\hline
$\alpha_i$&0&APE&HYP1&HYP2\\
$C^A_{\rm lat}$&26.26 &5.71&7.13&0.61\\
$C^S_{\rm lat}$&12.46&4.46&3.63&1.31\\
$\chi$&-6.90 &-0.54&-1.75&0.35\\
\hline
\end{tabular}
\end{center}
\caption{\label{tab3} Lattice contribution to the matching constant between the axial(scalar) static-light 
current regularised on the lattice and its counterpart renormalised in the continuum. We indicated the contribution
$\chi\equiv d_2 + h -q -2d^I$ coming from the chiral symmetry breaking term of the light quark action.}
\end{table}
\noindent We note that the one loop corrections for the set HYP2 are very small compared to the set $\alpha_i=0$, 
confirming the observation that UV fluctuations are strongly suppressed by this action \cite{Alpha2005}, which 
improves highly the signal/noise ratio. It is particularly impressive on the constant $e$ related to the static field 
renormalisation. In that case the tadpole contribution is much smaller for HYP2 than for Eichten-Hill (5.96 vs. 12.23)
and the "sunset" contribution is negative instead of positive (-9.58 vs. 12.25). Another interesting property of the HYP2 
action is that the contribution coming from the chiral symmetry breaking term of the light quark action is reduced 
compared to what is found with the 
other static quark actions, in particular HYP1, as indicated in the last row of Table $\ref{tab3}$. The main consequence 
is that the 
ratio $Z_V/Z_A$ between the matching constants of the vector and axial static-light currents is closer to 1.
Of course this feature is only true at one-loop of perturbation theory and can change at the non-perturbative level.

\section{$B_s - \bsb$ mixing with overlap fermions}

\noindent In this part we present the results of the computation of the renormalisation constants of static-light 
four-fermion 
operators with the light quark described by the Neuberger action. The bag parameter $B_{B_s}$ 
associated with the $B_s - \bsb$ mixing amplitude in the Standard Model is defined by
\bea\nonumber
B_{B_s}&=&\frac{\elematrice{\bsb}{(\bar{b}s)_{V-A}(\bar{b}s)_{V-A}}{B_s}}
{\elematrice{\bsb}{(\bar{b}s)_{V-A}(\bar{b}s)_{V-A}}{B_s}_{\rm VSA}},\\ 
\elematrice{\bsb}{(\bar{b}s)_{V-A}(\bar{b}s)_{V-A}}{B_s}_{\rm VSA}&=&
\elematrice{\bsb}{(\bar{b}s)_{V-A}}{0}\elematrice{0}{(\bar{b}s)_{V-A}}{B_s}\,.
\eea
We have to introduce in addition to the operator $O_1\equiv (\bar{b}s)_{V-A}(\bar{b}s)_{V-A}$ 
the following operators of the supersymmetric basis:
\bea\nonumber
O_2&=&(\bar{b}s)_{S-P}\; (\bar{b}s)_{S-P}\,,\\
\nonumber
O_3&=&(\bar{b}s)_{V-A}\; (\bar{b}s)_{V+A}\,,\\
O_4&=&(\bar{b}s)_{S-P}\; (\bar{b}s)_{S+P}\,.
\eea
Then we define as usual the bag parameters $B_{i=1,...,4}$ in terms of the Vacuum Saturation Approximation
matrix elements by
\beq\nonumber
\elematrice{\bsb}{O_i}{B_s}(\mu)=\elematrice{\bsb}{O_i}{B_s}_{\rm VSA}B_i(\m)\,.
\eeq
We define the HQET operators $\widetilde{O}_{i=1,...,4}$ by
\bea\nonumber
\widetilde{O}_1\equiv \widetilde{O}_{VV+AA}&=&(\bar{h}^{(+)}s)_{V-A}\;(\bar{h}^{(-)}s)_{V-A}\,,\\
\nonumber
\widetilde{O}_2\equiv \widetilde{O}_{SS+PP}&=&(\bar{h}^{(+)}s)_{S-P}\;(\bar{h}^{(-)}s)_{S-P}\,,\\
\nonumber
\widetilde{O}_3\equiv \widetilde{O}_{VV-AA}&=&(\bar{h}^{(+)}s)_{V-A}\;(\bar{h}^{(-)}s)_{V+A}\,,\\
\widetilde{O}_4\equiv \widetilde{O}_{SS-PP}&=&(\bar{h}^{(+)}s)_{S-P}\;(\bar{h}^{(-)}s)_{S+P}\,,
\eea
and their associated bag parameter $\widetilde{B}_i$, $i=1,2,3,4$.

\noindent The extraction of $B_{B_s}$ from our lattice simulation needs the following steps:\\
(1) $\tilde{B}^{\rm lat}_i(a)$ are matched onto the continuum $\overline{\rm MS}$(NDR) scheme at NLO in
perturbation theory at the renormalization scale $\mu=1/a$ \cite{becirevic},\\
(2) $\tilde{B}_i$ are evolved from $\mu=1/a$ to $\mu=m_b$ by using the HQET anomalous
dimension matrix, known to 2-loop accuracy in perturbation theory \cite{spqcdr,run},\\
(3) $\tilde{B}_i(\mu=m_b)$ are finally matched onto their QCD counterpart, $B_i(m_b)$, 
in the $\overline{\rm MS}$(NDR) scheme at NLO \cite{run}.\\
The matching scales are such that neither $\ln (a\mu)$ in step (1) nor 
$\ln(\mu/m_b)$ in step (3) correct strongly the matching constants. In the following we will 
concentrate on step (1).

\noindent The total lattice fermionic action is $S=S^{\rm HQET}+S^N_L$ where
\bea\nonumber
S^{\rm HQET}_H&=&a^3\sum_{n} 
\left\{\bar{h}^+(n)\left[h^+(n)-V^{\dag,{\rm HYP}}_4(n-\hat{4})h^+(n-\hat{4})\right]\right.\\
\nonumber
&&\left.\hspace{1cm}-
\bar{h}^-(n)\left[V^{{\rm HYP}}_4(n)h^-(n+\hat{4})-h^-(n)\right]\right.\\
\nonumber
&&\left.\hspace{1cm} + \,\delta m\, \left[\bar{h}^+(n)h^+(n)+\bar{h}^-(n)h^+(n)\right]\right\}\,,
\eea
\beq
S^{\rm N}_L=a^3 \sum_n \bar{\psi}(n) D_N(m_0)\,\psi(n)\,,\quad
D_N(m_0)=\left(1-\frac{1}{2\rho} am_0 \right)D_N+am_0\,,
\eeq
\beq\nonumber
D_N=\frac{\rho}{a}\left(1+\frac{X}{\sqrt{X^\dag X}}\right),\quad X=D_W -\frac{\rho}{a} 
\quad, 0<\rho<2\,.
\eeq
The static quark (antiquark) field satisfies the equation of motion
\beq\nonumber
\gamma_0 h^\pm(x) = \pm h^\pm(x).
\eeq
The HQET action is invariant under the finite Heavy Quark Symmetry (HQS) transformations
\beq
\bar{h}^{(\pm)}(x)\stackrel{HQS(i)}{\longrightarrow} 
-\frac 1 2 \epsilon^{ijk}\bar{h}^{(\pm)}(x)\gamma_j\gamma_k\quad  (i=1,2,3)\label{HQS}\,,
\eeq
and the overlap action is invariant under the infinitesimal chiral transformation \cite{chiralinvoverlap}
\beq
\psi \ra \left[1+i\epsilon \gamma^5\left(1-\frac{a}{2}D_N\right)\right]\psi, \quad \bar{\psi}\ra
\bar{\psi} \left[1+i\epsilon \left(1-\frac{a}{2}D_N\right)\gamma^5\right]\,.
\eeq
The matching between the operators regularised on the lattice and their counterpart of the continuum needs 
normally 16 matching constants,
as $\widetilde{O}_1$ and $\widetilde{O}_2$ can mix with
$\widetilde{O}_3$ and $\widetilde{O}_4$:
\beq\nonumber
\widetilde{O}^{\overline{\rm MS}}_i(\mu)=Z_{ij}(a\mu)\widetilde{O}_j(a),\quad i=1,...,4, \quad j=1,...,4\,.
\eeq
However, thanks to Heavy Quark Symmetry, these constants are not all independent. Here we give the
details of the proof, as
it was not fully presented in \cite{becirevic} or \cite{proceedingBB} (it was independently
presented and generalised in \cite{PPPW}). Under the HQS transformation (\ref{HQS}), one has
\beq\nonumber
\widetilde{O}_{SS+PP}\equiv-\widetilde{O}_{(VV+AA)_0}, \quad
\widetilde{O}_{VV+AA}\stackrel{HQS(i)}{\longrightarrow}\widetilde{O}_{VV+AA}, \quad 
\widetilde{O}_{SS+PP}\stackrel{HQS(i)}{\longrightarrow}-\widetilde{O}_{(VV+AA)_i}\,,
\eeq
\beq\nonumber
\widetilde{O}_{VV-AA}\stackrel{HQS(i)}{\longrightarrow}\sum_{j=1,3}^{j\neq i}
\widetilde{O}_{(VV-AA)_j}-(\widetilde{O}_{(VV-AA)_i}+\widetilde{O}_{(VV-AA)_0})
\equiv(\widetilde{O}_{VV-AA})_\perp - (\widetilde{O}_{VV-AA})_\parallel\,,
\eeq
\beq\nonumber
\widetilde{O}_{SS-PP}\equiv-\widetilde{O}_{(VV-AA)_0},\quad 
\widetilde{O}_{SS-PP}\stackrel{HQS(i)}{\longrightarrow}\widetilde{O}_{(VV-AA)_i}\,.
\eeq
The different constraints are the followings:
\bea\nonumber
\lgl \widetilde{O}_{VV+AA}(\mu) \rgl&=& Z_{11} \lgl \widetilde{O}_{VV+AA}(a) \rgl +Z_{12} \lgl \widetilde{O}_{SS+PP}(a) \rgl
+Z_{13} \lgl \widetilde{O}_{VV-AA}(a) \rgl +Z_{14} \lgl \widetilde{O}_{SS-PP}(a) \rgl\,,\\
\nonumber
\lgl \widetilde{O}_{VV+AA}(\mu) \rgl&=& Z_{11} \lgl \widetilde{O}_{VV+AA}(a) \rgl -Z_{12} \lgl \widetilde{O}_{(VV+AA)_i}(a) \rgl 
+ Z_{13} (\lgl \widetilde{O}_{VV-AA}(a) \rgl_\perp - \lgl \widetilde{O}_{VV-AA}(a) \rgl_\parallel)\\ 
\nonumber
&+&Z_{14} \lgl \widetilde{O}_{(VV-AA)_i}(a) \rgl\quad ({\rm HQS(i)})\,,
\eea
\bea\nonumber
\sum_{i=1,3}\lgl \widetilde{O}_{VV+AA}(\mu) \rgl&\equiv& 3 \lgl \widetilde{O}_{VV+AA}(\mu) \rgl\\
\nonumber
&=&(3Z_{11}-Z_{12}) \lgl \widetilde{O}_{VV+AA}(a) \rgl -Z_{12} \lgl \widetilde{O}_{SS+PP}(a)\rgl +
(Z_{13}+Z_{14})\lgl \widetilde{O}_{VV-AA}(a) \rgl\\
\nonumber
&+& (Z_{14}+4 Z_{13})\lgl \widetilde{O}_{SS-PP}(a) \rgl\,,
\eea
implying that
\beq\label{cont1}
Z_{12}=0, \quad Z_{14}=2 Z_{13}\,.
\eeq
\vspace{-0.2cm}
\bea\nonumber
\lgl \widetilde{O}_{SS+PP}(\mu) \rgl&=& Z_{21} \lgl \widetilde{O}_{VV+AA}(a) \rgl +Z_{22} \lgl \widetilde{O}_{SS+PP}(a) \rgl
+ Z_{23} \lgl \widetilde{O}_{VV-AA}(a) \rgl +Z_{24} \lgl \widetilde{O}_{SS-PP}(a)\rgl\,,\\
\nonumber
-\lgl \widetilde{O}_{(VV+AA)_i}(\mu) \rgl&=& Z_{21} \lgl \widetilde{O}_{VV+AA}(a) \rgl -Z_{22} \lgl \widetilde{O}_{(VV+AA)_i}(a) 
\rgl +Z_{23} (\lgl \widetilde{O}_{VV-AA}(a) \rgl_\perp - \lgl \widetilde{O}_{VV-AA}(a) \rgl_\parallel)\,\\ 
\nonumber
&+&Z_{24} \lgl \widetilde{O}_{(VV-AA)_i}(a) \rgl\quad ({\rm HQS(i)})\,,
\eea
\vspace{-0.8cm}
\bea\nonumber
-\sum_{i=1,3} \widetilde{O}_{(VV+AA)_i}(\mu) \pm \widetilde{O}_{(VV+AA)_0}(\mu) &\equiv& -\lgl \widetilde{O}_{SS+PP}(\mu) \rgl -
\lgl \widetilde{O}_{VV+AA}(\mu) \rgl\\ 
\nonumber
&=& (3Z_{21}-Z_{22}) \lgl \widetilde{O}_{VV+AA}(a) \rgl  - Z_{22}\lgl \widetilde{O}_{SS+PP}(a)\rgl\\
\nonumber
&+& (Z_{23}+Z_{24})\lgl \widetilde{O}_{VV-AA}(a) \rgl + (Z_{24}+4 Z_{23}) \lgl \widetilde{O}_{SS-PP}(a)\rgl\\
\nonumber
&=& -(Z_{11}+Z_{21}) \lgl \widetilde{O}_{VV+AA}(a) \rgl - Z_{22} \lgl \widetilde{O}_{SS+PP}(a)\rgl \\
\nonumber
&-&[ (Z_{13}+Z_{23})\lgl \widetilde{O}_{VV-AA}(a) \rgl +(Z_{14}+Z_{24})\lgl \widetilde{O}_{SS-PP}(a) \rgl]\,,
\eea
giving the constraints 
\beq\label{cont2}
Z_{21}=\frac{Z_{22}-Z_{11}}{4}, \quad Z_{24}=-(Z_{13}+2 Z_{23})\,.
\eeq
\vspace{-0.8cm}
\bea\nonumber
\lgl \widetilde{O}_{VV-AA}(\mu) \rgl&=& Z_{31} \lgl \widetilde{O}_{VV+AA}(a) \rgl +Z_{32} \lgl \widetilde{O}_{SS+PP}(a) \rgl
+Z_{33} \lgl \widetilde{O}_{VV-AA}(a) \rgl +Z_{34} \lgl \widetilde{O}_{SS-PP}(a) \rgl\,,\\
\nonumber
\lgl \widetilde{O}_{SS-PP}(\mu) \rgl&=& Z_{41} \lgl \widetilde{O}_{VV+AA}(a) \rgl +Z_{42} \lgl \widetilde{O}_{SS+PP}(a) \rgl 
+ Z_{43} \lgl \widetilde{O}_{VV-AA}(a) \rgl +Z_{44} \lgl \widetilde{O}_{SS-PP}(a) \rgl\,,\\
\nonumber
\lgl \widetilde{O}_{(VV-AA)_i}(\mu) \rgl&=& Z_{41} \lgl \widetilde{O}_{VV+AA}(a) \rgl -Z_{42} \lgl \widetilde{O}_{(VV+AA)_i}(a) 
\rgl +Z_{43} (\lgl \widetilde{O}_{VV-AA}(a) \rgl_\perp - \lgl \widetilde{O}_{VV-AA}(a) \rgl_\parallel )\\
\nonumber
&+&Z_{44} \lgl \widetilde{O}_{(VV-AA)_i}(a) \rgl\quad ({\rm HQS(i)})\,,
\eea
\bea\nonumber
\sum_{i=1,3} \widetilde{O}_{(VV-AA)_i}(\mu) \pm \widetilde{O}_{(VV-AA)_0}(\mu) &\equiv& \lgl \widetilde{O}_{SS-PP}(\mu) \rgl +
\lgl \widetilde{O}_{VV-AA}(\mu) \rgl\\ 
\nonumber
&=& (3Z_{41}-Z_{42}) \lgl \widetilde{O}_{VV+AA}(a) \rgl  - Z_{42}\lgl \widetilde{O}_{SS+PP}(a)\rgl\\
\nonumber
&+& (Z_{43}+Z_{44})\lgl \widetilde{O}_{VV-AA}(a) \rgl + (Z_{44}+4 Z_{43}) \lgl \widetilde{O}_{SS-PP}(a)\rgl\\
\nonumber
&=& (Z_{31}+Z_{41}) \lgl \widetilde{O}_{VV+AA}(a) \rgl + (Z_{32} + Z_{42} \lgl \widetilde{O}_{SS+PP}(a)\rgl \\
\nonumber
&+& (Z_{33}+Z_{43})\lgl \widetilde{O}_{VV-AA}(a) \rgl + (Z_{34}+Z_{44})\lgl \widetilde{O}_{SS-PP}(a) \rgl\,.
\eea
One obtains eventually the constraints 
\beq\label{cont3}
Z_{44}=Z_{33}, \quad Z_{42}=-\frac{Z_{32}}{2}, \quad
Z_{41}=\frac{2Z_{31}-Z_{32}}{4}, \quad Z_{43}=\frac{Z_{34}}{4}\,.
\eeq
The renormalisation matrix has the following structure:
\beq
Z=\left(
\begin{array}{cccc}
Z_{11}&0&Z_{13}&2 Z_{13}\\
\frac{Z_{22}-Z_{11}}{4}& Z_{22}& Z_{23} &-(Z_{13}+2 Z_{23})\\
Z_{31}&Z_{32}&Z_{33}&Z_{34}\\
\frac{2Z_{31}-Z_{32}}{4}&-\frac{Z_{32}}{2}&\frac{Z_{34}}{4}&Z_{33}\\
\end{array}
\right)\,.
\eeq
Further constraints are obtained thanks to the invariance of the overlap action under the finite chiral 
transformation
\beq\nonumber
\psi \ra i\gamma^5\left(1-\frac{a}{2}D_N\right)\psi, \quad \bar{\psi}\ra
i\bar{\psi} \left(1-\frac{a}{2}D_N\right)\gamma^5\,.
\eeq
Under such a transformation one has
\beq\nonumber
\widetilde{O}_{VV+AA}\to - \widetilde{O}_{VV+AA}, \quad \widetilde{O}_{SS+PP} \to - 
\widetilde{O}_{SS+PP}\,,
\eeq
\beq\nonumber
\widetilde{O}_{VV-AA}\to + \widetilde{O}_{VV-AA}, \quad \widetilde{O}_{SS-PP} \to + \widetilde{O}_{SS-PP}\,.
\eeq
The final result is then 
\beq\label{Z4f}
Z=\left(
\begin{array}{cccc}
Z_{11}&0&0&0\\
\frac{Z_{22}-Z_{11}}{4}& Z_{22}& 0&0\\
0&0&Z_{33}&Z_{34}\\
0&0&\frac{Z_{34}}{4}&Z_{33}\\
\end{array}
\right).
\eeq
There is no mixing of left-left four-fermion static-light operators regularised on the lattice with dimension 6 operators 
of different chirality, reducing significantly
the systematic error coming from such a spurious mixing when the light quark is described by the Wilson-Clover action: 
indeed the
matching of those operators with their counterpart renormalised in the continuum $\overline{\rm MS}$ scheme does not 
need any subtraction.

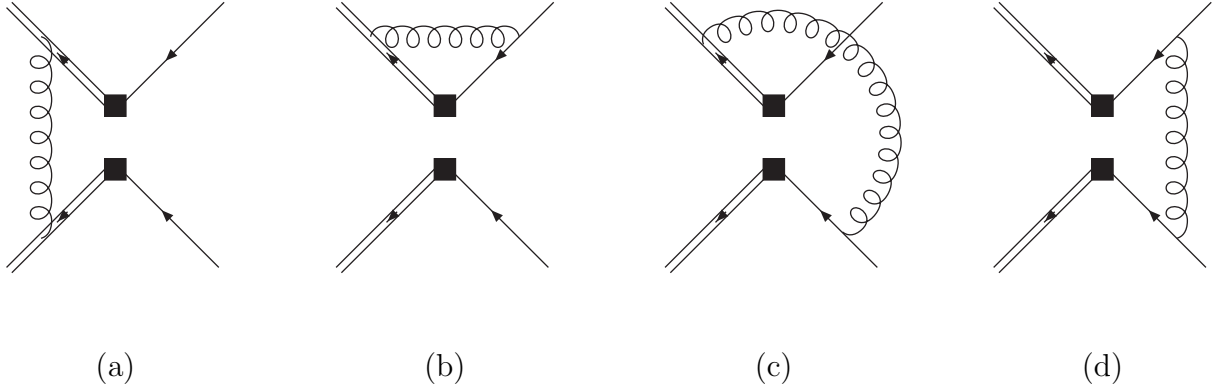
\begin{figure}[t]
\begin{center}
\begin{tabular}{ccccccc}
\begin{picture}(0,30)(0,0)
\Line(1,1)(-39,41)
\Gluon(-28,28)(-28,-48){4}{7}
\ArrowLine(41,41)(0,0)
\ArrowLine(-18,18)(-22,22)
\Line(-1,-1)(-41,39)
\CBox(-4,-2)(4,6){Black}{Black}
\CBox(-4,-26)(4,-18){Black}{Black}
\Line(-1,-19)(-41,-59)
\ArrowLine(39,-59)(-1,-19)
\ArrowLine(-18,-38)(-22,-42)
\Line(1,-21)(-39,-61)
\end{picture}
&\rule[0cm]{3.0cm}{0cm}&

\begin{picture}(0,30)(0,0)
\Line(1,1)(-39,41)
\ArrowLine(41,41)(0,0)
\ArrowLine(-18,18)(-22,22)
\Gluon(-28,28)(28,28){4}{6}
\Line(-1,-1)(-41,39)
\CBox(-4,-2)(4,6){Black}{Black}
\CBox(-4,-26)(4,-18){Black}{Black}
\Line(-1,-19)(-41,-59)
\ArrowLine(39,-59)(-1,-19)
\ArrowLine(-18,-38)(-22,-42)
\Line(1,-21)(-39,-61)
\end{picture}

&\rule[0cm]{3.0cm}{0cm}&
\begin{picture}(0,30)(0,0)
\Line(1,1)(-39,41)
\ArrowLine(41,41)(0,0)
\ArrowLine(-18,18)(-22,22)
\GlueArc(0,-10)(44,-54,127){4}{15}
\Line(-1,-1)(-41,39)
\CBox(-4,-2)(4,6){Black}{Black}
\CBox(-4,-26)(4,-18){Black}{Black}
\Line(-1,-19)(-41,-59)
\ArrowLine(39,-59)(-1,-19)
\ArrowLine(-18,-38)(-22,-42)
\Line(1,-21)(-39,-61)
\end{picture}

&\rule[0cm]{3.0cm}{0cm}&
\begin{picture}(0,30)(0,0)
\Line(1,1)(-39,41)
\ArrowLine(41,41)(0,0)
\ArrowLine(-18,18)(-22,22)
\Gluon(28,28)(28,-48){4}{7}
\Line(-1,-1)(-41,39)
\CBox(-4,-2)(4,6){Black}{Black}
\CBox(-4,-26)(4,-18){Black}{Black}
\Line(-1,-19)(-41,-59)
\ArrowLine(39,-59)(-1,-19)
\ArrowLine(-18,-38)(-22,-42)
\Line(1,-21)(-39,-61)
\end{picture}
\\
\vspace{2.5cm}
&&&&&&\\
(a)&&(b)&&(c)&&(d)\\
\end{tabular}
\end{center}
\caption{\label{figdiag4f} Diagrams giving the one loop correction to a static-light four-fermion operator.}
\end{figure}
\noindent We recall that the overlap propagator without mass reads\footnote{We invite the reader to have a look in 
Appendix~A in which 
the notations used in those equations are made more precise.}
\beq
S^{ab}_{\rm overlap}(k)=\delta^{ab}\frac{a}{2\rho}\left(\frac{-i \Ga}{\omega+b} + 1\right)\,, \quad
b(k) = W(k) - \rho\,, \quad \omega(k)=a\left(\sqrt{X^\dag X}\right)_0(k)\,,
\eeq
where $X_0$ is the free part of the Wilson kernel with a negative mass $-\frac{\rho}{a}$, 
and the quark-quark-gluon vertex is defined by \cite{vertexoverlap} 

\beq
V^{a, \rm{overlap}}_{\mu, qqg} (p,p') = -ig_0 T^a \, \frac{\rho}{\omega(p) + \omega(p')} 
\Bigg[ \gamma^\mu c_\mu -i s_\mu + \frac{a^2}{\omega(p)\omega(p')} X_0(p') 
\Bigg( \gamma^\mu c_\mu + i s_\mu \Bigg)X_0(p) \Bigg]\, .
\eeq
The renormalisation constants of dimension 6 static-light four-fermion operators are given at one loop of perturbation 
theory by the diagrams of Figure \ref{figdiag4f}.

\noindent Following the notations of \cite{becirevic}, the matching constants are defined by
\beqns
Z_{11}^{\msb}&=&1+\frac{\alpha^{\msb}_s}{4\pi}\left[\frac{7}{3}+\frac{d_s}{3}-\frac{10d_1}{3} - \frac{c}{3} 
-\frac{4e}{3} -\frac{4f}{3} + \frac{2d_{\xi}}{3}+4 \ln(a^2\mu^2)\right],\\
Z_{21}^{\msb}&=&\frac{\alpha^{\msb}_s}{4\pi}\left[-\frac{5}{36}-\frac{d_s}{36} -\frac{2d_v}{9} +\frac{d_1}{2} 
+\frac{c}{4}-  \frac{d_{\xi}}{6}-\frac{2}{3}\ln (a^2\mu^2)\right],\\
Z_{22}^{\msb}&=&1+\frac{\alpha^{\msb}_s}{4\pi}\left[\frac{16}{9}+\frac{2d_s}{9}-\frac{8d_v}{9} -\frac{4d_1}{3} 
+\frac{2c}{3}-\frac{4e}{3}-\frac{4f}{3}+ \frac{4}{3} \ln(a^2\mu^2)\right],\\
Z_{33}^{\msb}&=&1+\frac{\alpha^{\msb}_s}{4\pi}\left[\frac{41}{12}-\frac{d_v}{6} - \frac{7d_1}{3} + \frac{c}{6}
- \frac{4e}{3}-\frac{4f}{3}+\frac{7d_{\xi}}{6}+\frac{7}{2}\ln(a^2\mu^2)\right],\\
Z_{34}^{\msb}&=&\frac{\alpha^{\msb}_s}{4\pi}\left[\frac{1}{2}-d_v +\, 2d_1+c -d_\xi  
- 3\ln(a^2\mu^2)\right],\\
Z_{43}^{\msb}&=&\frac{\alpha^{\msb}_s}{4\pi}\left[\frac{1}{8}-\frac{d_v}{4} +\frac{d_1}{2}+\frac{c}{4}
-\frac{d_\xi}{4}-\frac{3}{4}\ln(a^2\mu^2)\right],\\
Z_{44}^{\msb}&=&1+\frac{\alpha^{\msb}_s}{4\pi}\left[\frac{41}{12}-\frac{d_v}{6} -\frac{7d_1}{3}+\frac{c}{6}
- \frac{4e}{3}-\frac{4f}{3}+\frac{7d_\xi}{6}+\frac{7}{2}\ln(a^2\mu^2)\right],\\
\eeqns
where $c$ and $d_1$ correspond to diagrams \ref{figdiag4f}(a) and \ref{figdiag4f}(b) respectively.
The matching constant of the axial static-light current is defined by
\beq
Z^{\msb}_A=1+\frac{\alpha_s}{12\pi^2}\left[\frac{5}{4}-\frac{e+f}{2}-d_1
+\frac{3}{2} \ln (a^2\m^2)\right]. 
\eeq
We have collected the numerical values of $c$ and $d_1$ in Table \ref{tab5} and we have given their analytical 
expression in Appendix~C. We agree with the authors of \cite{becirevic} for the analytical expression of 
$d_1(\alpha_i = 0)$ \cite{becirevicprivate} and for its numerical value.
 $f(\rho)$, $d_s(\rho)$ and $d_v(\rho)$, involving only light quark legs and computed in \cite{selfenergyoverlap},  
are included in the same table for $\rho=1.4$ and 1.6 that we chose to perform the lattice simulation, and 
$d_\xi=-4.792010$.
We obtain for $\rho=1.4$ and the set HYP1
\beq
\begin{array}{ll}
Z^{\msb}_{11}(1/a)=1+\frac{\alpha^{\msb}_s(1/a)}{4\pi}\times 20.0579\,,&
Z^{\msb}_{22}(1/a)=1+\frac{\alpha^{\msb}_s(1/a)}{4\pi}\times 19.6915\,,\\
Z^{\msb}_A(1/a)=1+\frac{\alpha^{\msb}_s(1/a)}{4\pi}\times 11.2557\,.&\\
\end{array}
\eeq
Here we would like to make two remarks.\\ 
The first one is that the bag parameters $\widetilde{B}^{\msb}(\mu)_i$ are matched to $\widetilde{B}(1/a)_i$ with 
$\frac{Z_{ij}}{Z^2_A}$: in the ratio the quark self-energies cancel, reducing the corrections.\\
The second remark concerns the numerical value of the renormalisation constants: one needs to define the expansion
parameter $\alpha_s$ in terms of the lattice coupling, in order to improve as much as possible the 
perturbative computation. We decided in our analysis to use the constant $\alpha^V(3.41/a)$, that is related to the average 
plaquette $\lgl 1/3 
{\rm Tr} (U_{\square})\rgl$ \cite{LepageMackenzie}, and the ratio $\Lambda_{\msb}/\Lambda_V$, to compute $\alpha^{\msb}_s(1/a)$ 
at two loops of perturbation theory. An alternative approach could have been to choose the scale $\mu = q^*$ between 
$1/a$ and $\pi/a$, as done in \cite{gimenez}, and include the spreading in the systematic error as done in \cite{spqcdr}.
Of course in that case the logarithmic terms appearing in (\ref{matching}) must be taken into account.
\begin{table}[b]
\begin{center}
\begin{tabular}{ccc}
\begin{tabular}{|c|c|c|}
\hline
$\rho$&1.4&1.6\\
\hline
$f(\rho)$&-17.47&-13.24\\
$d_s(\rho)$&2.55&3.06\\
$d_v(\rho)$&0.056&0.068\\
$d_1(\rho,\alpha_i=0)$&0.648&0.707\\
$d_1(\rho,{\rm APE})$&0.320&0.346\\
$d_1(\rho,{\rm HYP1})$&0.285&0.306\\
$d_1(\rho,{\rm HYP2})$&0.032&0.026\\
\hline
\end{tabular}&&
\begin{tabular}{|c|c|c|c|c|}
\hline
$\alpha_i$&0&APE&HYP1&HYP2\\
\hline
$c$&4.53&-3.63&-3.24&-7.82\\
\hline
\end{tabular}
\end{tabular}
\end{center}
\caption{\label{tab5} Numerical values of $c$, $d_1(\rho)$, $f(\rho)$, $d_s(\rho)$ and $d_v(\rho)$ 
defined in the text.}
\end{table}

\section{Conclusion}\label{sec5}

In this paper we have calculated the one loop corrections at $O(a)$ of static-light currents 
$\bar{h}\Gamma q$ and four-fermion operators $(\bar{h}\Gamma q) \, (\bar{h}\Gamma q)$ in lattice HQET 
with a hypercubic blocking of the Wilson line which defines the static quark propagator. 
It determines the renormalization of the operators which are used
to compute in the static limit of HQET the decay constant $f_B$ and the bag parameters $B_i$ 
associated with
the $B_s - \bsb$ mixing amplitude in the Standard Model and beyond. \\
In particular we have given values of the renormalisation constants of the static-light four-fermion
operators when the light quark is described by the overlap action, which is an elegant way to restore on the lattice the chiral 
symmetry of the continuum but is highly demanding in computation time, so that a non perturbative renormalisation 
procedure, like the Schr\"odinger functional scheme \cite{schrodscheme}, is not underway yet. However a further step 
could be to compute in this scheme -- i.e. non perturbatively -- the matching constants of static-light bilinear 
currents 
when the light quark is described in the bulk by the Neuberger operator \cite{neubergerluscher}.

\section*{Acknowledgment}

\noindent I gratefully acknowledge helpful discussions with D. Be\'cirevi\'c, N. Garron, 
A. Le Yaouanc, C. Michael, A. Shindler 
and R. Sommer. This work is supported in part by the EU Contract 
No.~MRTN-CT-2006-035482 (``FLAVIAnet'') and by the Deutsche Forschungsgemeinschaft in the
SFB/TR~09.

\appendix

\section{Notations}

We give here the notations that appear in the main part of the paper and below in the analytical expressions of 
matching constants.

\beq\nonumber
\int_k \equiv \int_{-\pi}^{\pi} \frac{d^4k}{(2\pi)^4}\;,\quad\quad
\int_{\vec{k}} \equiv \int_{-\pi}^{\pi} \frac{d^3k}{(2\pi)^3}\;,
\eeq
\beq\nonumber
U_\mu(n)=e^{iag_0A^a_\mu(n)T^a}=1+iag_0A^a_\mu(n)T^a-\frac{a^2g_0^2}{2!}A^a_\mu(n)A^b_\mu(n)T^aT^b
+{\cal O}(g^3_0),
\eeq
\beq\nonumber
U^{\rm HYP}_\mu(n)=e^{iag_0B^a_\mu(n)T^a}=1+iag_0B^a_\mu(n)T^a-\frac{a^2g^2}{2!}B^a_\mu(n)B^b_\mu(n)T^aT^b
+{\cal O}(g^3_0),
\eeq
\beq\nonumber
A^a_\mu(n)=\int_p e^{ip(n+\frac{a}{2})}A^a_\mu(p), \quad\quad
B^a_\mu(n)=\int_p e^{ip(n+\frac{a}{2})}B^a_\mu(p),
\eeq
\beq\nonumber
F^2=\sum_{i=1}^4 F^2_i, \quad \vec{F}^2=\sum_{i=1}^3 F^2_i
\eeq
\beq\nonumber
\Gamma_{\lambda}=\sin ak_{\lambda}, \quad \quad
c_\mu=\cos \left(\frac{a(p+p')_\mu}{2}\right), \quad\quad
s_\mu=\sin \left(\frac{a(p+p')_\mu}{2}\right),
\eeq
\beq\nonumber
M_\mu=\cos \left(\frac{k_\mu}{2}\right), \quad\quad
N_\mu=\sin \left(\frac{k_\mu}{2}\right),
\quad\quad W=2 N^2,
\eeq
\beq\nonumber
E^2=\vec{N}^2 + \frac{a^2\lambda^2}{4}, \quad 
E^2_1=\frac{(\vec{N}^2)^2 + \frac{\vec{\Gamma}^2}{4}}{1+2\vec{N}^2}.
\eeq
\\
\beq\nonumber
B^{(1)}_\mu(k)=\sum_{\nu} h_{\mu\nu}(k)A_\nu(k),\quad
h_{\mu\nu}(k)=\delta_{\mu\nu}D_\mu(k) + (1-\delta_{\mu\nu})G_{\mu\nu}(k),
\eeq
\beq\nonumber
D_\mu(k)=1-c_1\sum_{\rho \neq \mu} N^2_\rho +c_2 \sum_{\rho < \sigma, \rho,\sigma \neq \mu}
N^2_\rho N^2_\sigma -c_3 N^2_\rho N^2_\sigma N^2_\tau,
\eeq
\beq\nonumber
G_{\mu\nu}(k)=N_\mu N_\nu \left(c_1-c_2 \frac{N^2_\rho + N^2_\sigma}{2} + c_3 
\frac{N^2_\rho N^2_\sigma}{3}\right) \equiv N_\mu N_\nu A'_\nu,
\eeq
\beq\nonumber
c_1=(2/3)\alpha_1[1+\alpha_2(1+\alpha_3)], \quad\quad c_2=(4/3) \alpha_1\alpha_2(1+2\alpha_3),
\quad\quad c_3=8\alpha_1\alpha_2\alpha_3.
\eeq

\section{Matching constants of ${\cal O}(a)$ improved operators}

Here we give the analytical expressions of the constants $d_1$, $d_2$, $d^I$, $n$, $h$ and $q$.

\bea\nonumber
d_1(\alpha_i)&=&\ln(a^2\lambda^2) 
+ (4\pi)^2 \left\{\frac{1}{16}\int_{\vec{k}}\frac{1}{1+2\vec{N}^2}\frac{1}{\sqrt{1+E^2_1}}
\frac{1}{E_1}\left(D_4 + \sum_{j=1}^3 A'_j N^2_j\right)\right.\\
\nonumber
&+&\left.\frac{1}{16}\int_{\vec{k}}\frac{1}{1+2\vec{N}^2}\frac{1}{E^2_1-E^2}
\left[D_4\left(\frac{\sqrt{1+E^2}}{E}-\frac{\sqrt{1+E^2_1}}{E_1}\right)\right.\right.\\
&&\hspace{3cm}\left.\left.+\sum_{j=1}^3 A'_j N^2_j M^2_j\left(\frac{1}{E\sqrt{1+E^2}}
-\frac{1}{E_1\sqrt{1+E^2_1}}\right)\right]\right\}\,\label{d1}\, ,
\eea
\beq
d_2(\alpha_i)=-\frac{1}{16}\int_{\vec{k}}\frac{D_4}{1+2\vec{N}^2}\frac{1}{E^2_1}, 
\quad d^I=-\int_{\vec{k}} \frac{D_4}{64}\frac{\vec{\Gamma}^2}{1+2\vec{N}^2} 
\frac{1}{\vec{N}^2 E^2_1}\label{d2dI}\, ,
\eeq
\bea\nonumber
n&=&\int_{\vec{k}}\frac{1}{16}\frac{1}{1+2\vec{N}^2}\left[\frac{D_4 \sqrt{1+E^2_1}}{E_1} +
\frac{\sum_j A'_j N^2_jM^2_j}{E_1\sqrt{1+E^2_1}}\right.\\
&&\hspace{1.4cm}\left. 
+\frac{D_4+\sum_j A'_j N^2_j}{4(E^2_1-\vec{N}^2)}
\left(\frac{4\vec{N}^2(1+\vec{N}^2)-\vec{\Gamma}^2}{\sqrt{\vec{N}^2}\sqrt{1+\vec{N}^2}}-
\frac{4E^2_1(1+E^2_1)-\vec{\Gamma}^2}{E_1\sqrt{1+E^2_1}}\right)\right]\, ,
\eea
\beq
h=-\int_{\vec{k}} \frac{1}{16} \frac{D_4}{\vec{N}^2},
\quad q=-\int_{\vec{k}} \frac{1}{64}\frac{D_4
\vec{\Gamma}^2}{1+2\vec{N}^2}\frac{1}{E^2_1}\label{hq}\, .
\eeq

\section{Static-light vertex with the overlap action}

Here we give the analytical expressions of $c$ and $d_1(\rho)$:

\bea
c(\alpha_i)&=&2\ln(a^2\lambda^2)+(4 \pi)^2\int_{\vec{k}} \frac{D^2_4 - E^2\sum^3_{i=1} 
N^2_i A'^2_i}{4E^3}\frac{1}{\sqrt{1+E^2}}\,,
\eea

\bea\nonumber
d_1(\rho)-\ln(a^2\lambda^2) - d_\xi&=&-(4\pi)^2\int_k \frac{\sum_j h_{4j}}{2iN_4 + \epsilon M_4} 
\frac{1}{2W+a^2\lambda^2}
\frac{-i\Ga + \omega+b}{2\rho(\omega+b)}\frac{\rho}{\omega+\rho}\\
\nonumber
&&\hspace{2cm}\left[\gamma_jM_j -i N_j
-\frac{i\Ga + b}{\omega}(\gamma_j M_j+i N_j)\right]\\
\nonumber
&=&(4\pi)^2 \int_k \frac{1}{2W+a^2\lambda^2}\frac{1}{\omega+b}
\frac{1}{\omega+\rho} \left[D_4\left(M^2_4 + \frac{\omega + b}{2}\right)\right.\\
&&\hspace{4cm}\left.+  
\sum_j A'_jN^2_j\left(M^2_j+\frac{\omega+b}{2}\right)\right].
\eea

\end{document}